\documentclass[twocolumn,showpacs]{revtex4}

\usepackage{epsfig}
\usepackage{amsmath}
\usepackage{amsfonts}
\usepackage{hyperref}
\usepackage{amssymb}
\usepackage{graphicx}

\def\Xint#1{\mathchoice
{\XXint\displaystyle\textstyle{#1}}%
{\XXint\textstyle\scriptstyle{#1}}%
{\XXint\scriptstyle\scriptscriptstyle{#1}}%
{\XXint\scriptscriptstyle\scriptscriptstyle{#1}}%
\!\int}
\def\XXint#1#2#3{{\setbox0=\hbox{$#1{#2#3}{\int}$}
\vcenter{\hbox{$#2#3$}}\kern-.5\wd0}}

\def\dashint{\Xint-}
\begin{document}
\title{{\bf Realistic Exact Solution for the Exterior Field of a Rotating Neutron Star}}
\author{Leonardo. A. Pach\'{o}n}
\email{lpachon@laft.org} \affiliation{Escuela de F\'isica,
Universidad Industrial de Santander, A.A. 678, Bucaramanga,
Colombia} \affiliation{Departamento de F\'isica, Universidad
Nacional, Bogot\'a D.C., Colombia}

\author{Jorge A. Rueda}
\email{jrueda@ula.ve} \affiliation{Escuela de F\'isica, Universidad
Industrial de Santander, A.A. 678, Bucaramanga, Colombia}
\affiliation{Centro de F\'isica Te\'orica, Universidad de Los Andes,
M\'erida, M\'erida 5101, Venezuela}

\author{Jos\'e D. Sanabria-G\'{o}mez}
\email{jsanabri@uis.edu.co} \affiliation{Escuela de F\'isica,
Universidad Industrial de Santander, A.A. 678, Bucaramanga,
Colombia}

\date{\today}

\begin{abstract}
A new six-parametric, axisymmetric and asymptotically flat exact solution of Einstein-Maxwell field
equations having reflection symmetry is presented. It has arbitrary physical parameters of mass,
angular momentum, mass--quadrupole moment, current octupole moment, electric charge and magnetic
dipole, so it can represent the exterior field of a rotating, deformed, magnetized and charged
object; some properties of the closed-form analytic solution such as its multipolar structure,
electromagnetic fields and singularities are also presented.  In the vacuum case, this analytic
solution is matched to some numerical interior solutions representing neutron stars, calculated by
Berti \& Stergioulas \cite{BertiStergioulas}, imposing that the multipole moments be the same. As an
independent test of accuracy of the solution to describe exterior fields of neutron stars, we
present an extensive comparison of the radii of innermost stable circular orbits (ISCOs) obtained
from Berti \& Stergioulas numerical solutions, Kerr solution \cite{Kerr}, Hartle \& Thorne solution \cite{HartleThorne}, an
analytic series expansion derived by Shibata \& Sasaki \cite{Shibata} and, our exact solution. We found
that radii of ISCOs from our solution fits better than others with realistic numerical interior
solutions.
\end{abstract}


\maketitle
%%%%%%%%%%%%%%%%%%%%%%%%%%%%%
\section{Introduction}
\label{Introduction}%%%%%%%%%
%%%%%%%%%%%%%%%%%%%%%%%%%%%%%

Observed pulsars are assumed to be highly deformed objects due to rotation. The spin frequencies of
the 11 nuclear-powered pulsars lie between 230 Hz (e.g. for the PSR 1845-19 pulsar) and 641--Hz
(e.g. for the PSR B1937+21 pulsar \cite{BackerKulkarni}, \cite{716Hz}), and within that range are
marginally consistent with a uniform distribution (see \cite{Michel}, \cite{Chakrabartyetal} and
references therein). The fast rotation is very important, not only in newly-born neutron stars
which may undergo secular and dynamical instabilities (see \cite{BertiBruni} and references
therein) but also in strange stars.

Studying Neutron Stars (NS) is interesting for several reasons. The quasi--stationary evolution of
an isolated NS can be tracked considering equilibrium sequences in which the rest mass is
constant while the angular momentum varies \cite{Cook94}. Such evolution may be driven by the
adiabatic loss of energy and angular momentum via electromagnetic or gravitational radiation. On
the other hand, the axisymmetric pulsations of rotating NS can be excited in several scenarios,
like core collapse, crust -- and core -- quakes or binary mergers, and could become detectable
in either gravitational waves or high-energy radiation \cite{FontStergioulas}. Furthermore, the
observational detection of such pulsations will yield valuable information about the equation of
state of relativistic stars and therefore information about the properties and behavior of the
matter at extreme conditions of high density ($\sim 10^{15}$ ${\rm gr/cm}^3$) and temperatures
(around $10^6$K) \cite{WiebickeGeppert}. In addition, new aspects of rotating NS have been revealed
in about 1000 pulsars. Some X-ray pulsars and some $\gamma$-ray pulsars have been detected in the
past years. Among these new objects, some exhibit quite different behaviour in their pulse periods
\cite{Kouveliotou}. The measurement of the period and its time derivative yields evidence of
ultra-magnetized stars, po\-ssibly representing magnetars \cite{DuncanThompson}.

In the last two decades, great advances have been made to understand
the properties of astrophysical objects like White Dwarfs, Black
Holes or NSs, in the frame of General Relativity. Particularly, from
the numerical approach, several physical properties of NS such as
energy density, inertia moments, equatorial radii and others are
known through the numerical solution of the full Einstein equations,
assuming different equations of state for NS and applying
algorithms, specially the Komatsu--Eriguchi--Hashisu self consistent
field--method \cite{KomatsuEriguchi} and its variants (e.g. in 2005
by Ghezzi \cite{Ghezzi}, Gusakov, Yakovlev \& Gnedin \cite{Gusakov}
and Berti,  White, Maniopoulou \& Bruni \cite{BertiBruni}. In 2004
by Berti \& Stergioulas \cite{BertiStergioulas} and Yoshida \&
Eriguchi \cite{YoshidaEriguchi}. In previous years by Stute \&
Camenzind in 2003 \cite{StuteCamenzind}, Manko, Mielke \&
Sanabria-G\'omez in 2000 \cite{SlowlyJD}, Sibgatullin \& Sunyaev
\cite{Sibgatullin} in 1998, Bocquet Bonazzola, Gourgoulhon \& Novak
\cite{BonazzolaBocquet}, in 1995, and by Cook, Shapiro \& Teukolsky
in 1994 \cite{Cook94} and 1992 \cite{Cook92}. and references
therein). On the other hand, from the analytic close approach, to
construct axisymmetric stationary exact solutions of the
Einstein--Maxwell equations with physical parameters defined from
the beginning is not a problem now, because of the development of
the powerful integral method by Sibgatullin (see
\cite{LibroSibgatullin} and \cite{MetodoSibgatullin}), which allows
to do this by considering the choosing of the Ernst's potentials on
the symmetry axis.

The aim of this paper is to present a new stationary axially
symmetric exact solution to the Einstein-Maxwell system as a model
for the exterior field of a NS, not only for rapidly but also for
slowly rotation rates. A consistent analytic closed representation
of the exterior space--time around a rapidly rotating NS is
desirable for many reasons, among them: (i) if we have an analytic
closed form for the me\-tric, the computation of the stationary
properties of the space--time is less difficult (for example,
geodesics in the exterior of NS could be studied analytically
\cite{Abramowicz} \cite{SibgatullinDG}; one could also find closed
form expressions for the radii and frequencies of the innermost
stable circular orbits (ISCOs) \cite{Abramowicz} \cite{Shibata},
etc.), (ii) it could be useful for studying the dynamical properties
of the space--time, like gravitational wave emission and chaotic
trajectories of particles around it. Furthermore, having an analytic
solution could simplify the calculation of properties of accretion
disks, epicyclic frequencies \cite{Sibgatullin} and so on.

Some analytic models for NS have been cons\-tructed and studied. Since the seminal works of Hartle
\cite{Hartle} and Hartle-Thorne \cite{HartleThorne}, the analytic structure of the space--time
outside a slowly rotating NS was generally associated with the Kerr solution because of its
simplicity and accuracy with numerical data obtained from numerical interior solutions. Nevertheless, Kerr solution does not present an accurate fit with the expected values of some of the physical properties of realistic stars in the regime of rapid rotation rates (like the radii of ISCOs, see \cite{Cook94, StuteCamenzind, BertiStergioulas} and the
mass--quadrupole moment \cite{Laarakkers}). Sibgatullin \& Sunyaev \cite{Sibgatullin}
adjusted the free parameters of the exact solution
by Manko, Martin, Ru\'iz, Sibgatullin \& Zaripov in \cite{MankoZaripov} with the numerical data for
the ISCO and the gravitational redshift of Cook, Shapiro \& Teukolsky \cite{Cook94}.
They analyzed the effects of the mass--quadrupole moment of a rapidly rotating NS on the energy
release in the equatorial layer on the surface of the accreting star and in the accretion disk,
using the normal sequences of Equations of State (EOS) A \cite{EOSA}, AU \cite{EOSAU} and FPS
\cite{EOSFPS}.

In 2000, Manko, Mielke \& Sanabria-G\'omez \cite{SlowlyJD} derived the charged, magnetized
generalization with arbitrary quadripolar deformation of the Tomimatsu-Sato $\delta=2$ solution
\cite{TSD2} written in rational functions, and claimed this solution as a possible model to describe the exterior
field of a rotating NS. Then, Stute \& Camenzind \cite{StuteCamenzind} decided to study that
solution to describe the gravitational field of a rapidly rotating NS. They
matched the solution by Manko {\it et al.} with the numerical interior
solutions from Cook {\it et al} \cite{Cook94} by a similar procedure to the one used in
\cite{Sibgatullin}, concluding that the accuracy of that solution is high in the case of the
redshift, but poor describing the radius of the marginally stable orbit (the matching was made
using non invariants local properties of the solution, as it was pointed out in
\cite{BertiStergioulas}). However, they considered that Manko {\it et al.} solution was a great
improvement compared with Hartle \& Thorne solution in the regime of fast rotation.

Recently, the full Einstein equations were solved in a numerical approach by Berti \& Stergioulas (B\&S)
\cite{BertiStergioulas}, to determine the NS space--time along of sequences of constant rest mass
for selected EOS (denoted as EOS A \cite{EOSA}, EOS AU \cite{EOSAU}, EOS FPS
\cite{EOSFPS}, EOS L \cite{EOSL} and EOS APRb \cite{EOSAPRb}). They matched the Manko {\it et. al.} solution
\cite{SlowlyJD} to the numerical solutions imposing the condition that the mass--quadrupole moment of the
numerical and analytic space--times be the same, concluding that the matching condition can be
satisfied only for very rapidly rotating stars. This affirmation was based on the fact that this exact
solution does not reduce conti\-nuously to the Schwarzschild
one when the rotation va\-nishes and, according to \cite{BertiStergioulas}, the mass--quadrupole
moment in the non--rotating limit (angular moment equals to zero) is very large for this
limit, i.e., $Q=-m(m^2+b^2)^2/(4(m^2-b^2))\, ,$ where $m$ is the mass and $b$ is a parameter
related with the arbitrary mass--quadrupole moment respectively. For that reason, it is not
possible that the solution by Manko {\it et al.} can describe NSs in the regime of slow
rotations. Besides, B\&S suggested that for inter\-media\-te rotation rates could
be used the exterior approximated analytic solution by Hartle \& Thorne \cite{HartleThorne}, valid
to second order in the rotation rate.

Following their own suggestion, in 2005 Berti {\it et al.} \cite{BertiBruni} compared
the Hartle-Thorne \cite{HartleThorne} slow-rotation approximation keeping terms up to second order
in the stellar angular velocity, and Manko {\it et al.} solution \cite{SlowlyJD} again with their
numerical solutions to the full Einstein's equations. Using the same matching procedure presented in
\cite{BertiStergioulas}, they found that the Hartle-Thorne appro\-xi\-mation offers very good
predictions for the corotating $R_+$ and counterrotantig $R_-$ ISCO radii, with $R_\pm$ accurate
better than 1\% even for the fastest millisecond pulsars. At these rotational rates the accuracy
on the mass--quadrupole moment is $\sim$20\%, and better for longer periods.

In spite of the fact that the rich phenomenology observed from pulsars
motivates the study of the interior and exterior electromagnetic fields of
rotating NS, we use the same procedure exposed in \cite{BertiStergioulas}, i.e., the matching of the
multipole moments in the vacuum case, because of the limited
information available about electromagnetic properties in numerical interior
solutions to Einstein­-Maxwell system. However, considering that the study of
electromagnetic fields in strongly curved space--times has been the
subject of past and recent interest, we also present the closed form and some properties
of the electromagnetic field of our solution.

This paper is presented in the following order. In section 2, we present the six-parametric
solution including issues about its construction. In section 3, we summarize the properties of this
solution through its multipolar structure, limiting cases, electromagnetic fields and
singularities. In section 4, we match our analytic solution with the interior numerical solutions by
Berti \& Stergioulas (hereafter B\&S) and present comparisons of the radii of ISCOs of several numerical and
exact solutions. Finally, concluding remarks about the paper are written.

%%%%%%%%%%%%%%%%%%%%%%%%%%%%%%%%%%%%%%
\section{Six-Parameter Solution}
\label{analgravfield}%%%%%%%%%%%%%%%%%
%%%%%%%%%%%%%%%%%%%%%%%%%%%%%%%%%%%%%%

In order to construct a realistic exact solution to the Einstein--Maxwell equations,
we must consider some physical assumptions.  In
Newtonian Theory of Gravitation all equilibrium states of isolated self-gravitating fluids (i.e.
stellar models) must have reflection symmetry through a plane which
is perpendicular to the rotation axis of the star \cite{Lichtenstein}; according to that idea, it has been
conjectured that stationary general relativistic stellar models must have reflection symmetry as
well \cite{Lindblom}.  For the stationary axisymmetric case a simple form of the metric was given
by Papapetrou \cite{Papap}, that is
\begin{equation}
\label{Papapetrou} ds^2=-f(dt-\omega d\phi)^2+f^{-1}\left[e^{2\gamma} (d\rho^2+dz^2)+\rho^2d\phi^2
\right]\, ,
\end{equation}
where $f$, $\omega$ and $\gamma$ are functions of the
quasi--cylindrical Weyl coordinates $(\rho,z)$.

Using the above line element, the Einstein-Maxwell equations can be reformulated, via Ernst's
procedure \cite{Ernst2}, in terms of two complex potentials ${\cal E}$ and $\Phi$ as follow:
\begin{eqnarray}
({\rm Re}\,{\cal E}+|\Phi|^2)\nabla^2{\cal E}&=& (\nabla {\cal E} +
2\Phi^*\nabla\Phi)\cdot\nabla{\cal E}, \nonumber\\ ({\rm Re}\,{\cal E}+|\Phi|^2)\nabla^2\Phi&=&
(\nabla {\cal E} + 2\Phi^*\nabla\Phi)\cdot\nabla\Phi\, . \label{Ernst}
\end{eqnarray}

In this paper, we solve the Ernst's equations (\ref{Ernst}) with the
aid of the Sibgatullin's method \cite{MetodoSibgatullin}
\cite{LibroSibgatullin}, according to which the complex potentials
$\cal E$ and $\Phi$ can be calculated from specified axis data
$e(z):={\cal E}(z,\rho=0)$ and \mbox{$f(z):=\Phi(z,\rho=0)$}, by the
integrals
\begin{eqnarray}
{\cal E}(z,\rho)&=&\frac{1}{\pi}\int_{-1}^1 \frac{e(\xi)\mu(\sigma)d\sigma}{\sqrt{1-\sigma^2}},
\nonumber\\
\label{Ernst1} \Phi(z,\rho)&=&\frac{1}{\pi}\int_{-1}^1
\frac{f(\xi)\mu(\sigma)d\sigma}{\sqrt{1-\sigma^2}}\, .
\end{eqnarray}

The unknown function $\mu(\sigma)$ must satisfy the singular integral equation
\begin{equation}
\dashint_{-1}^{1}\frac{\mu(\sigma)[e(\xi)+\tilde
e(\eta)+2f(\xi)\tilde
f(\eta)]d\sigma}{(\sigma-\tau)\sqrt{1-\sigma^2}}=0
\end{equation}
and the normalizing condition
\begin{equation}
\int_{-1}^1\frac{\mu(\sigma)d\sigma}{\sqrt{1-\sigma^2}}=\pi,
\end{equation}
where $\xi=z+i\rho\sigma$, $\eta=z+i\rho\tau$, $\rho$ and $z$ being the Weyl-Papapetrou
quasi--cylindrical coordinates, $\sigma, \tau\in[-1,1]$, $\tilde e(\eta):=\overline{e(\bar\eta)}$,
$\tilde f(\eta):=\overline{f(\bar\eta)}$ and the overbar stands for complex conjugation.

With the purpose to construct reflectionally symmetric solutions, the Ernst's potentials on
the symmetry axis $e(z)$ and $f(z)$ must satisfy the following conditions
\begin{equation}\label{condef}
e(z){\bar e}(-z)=1\, \quad {\rm and} \quad f(z)=-{\bar f}(-z)e(z)\, ,
\end{equation}
when the electromagnetic potentials are even functions of $z$
\cite{Simetria}. One choice of Ernst's potentials satisfying the
above considerations is  \cite{ez}
\begin{eqnarray}
e(z) = \frac{z^3-z^2(m+ia)-kz+is}{z^3+z^2(m-ia)-kz+is}\, ,
\nonumber\\\label{Potenciales eje} f(z) = \frac{q z^2+i\mu
z}{z^3+z^2(m-ia)-kz+is}\, ,
\end{eqnarray}
where $m$ denotes the gravitational mass of the source, $a$ its specific angular momentum, $q$
its electric charge, and $k$, $s$ and $\mu$ parameters related with the mass--quadrupole moment,
the current octupole and the dipolar magnetic moment respectively (see Subsection \ref{subsec:multipoles}).
We have included the parameter $s$
following the suggestion given in \cite{StuteCamenzind}
regarding to the Manko {\it et al.} solution,
whose accuracy is not good for the radius of the
marginally-stable-orbit, which is more sensitive to higher multipole
moments of the space-time. It seems to show that
higher multipole moments ought to be included, especially for rapid
rotation and stiffer equation of state, to improve the freedom to fit the
exterior gravitational field of neutron stars with numerical interior solutions. Moreover, this
parameter could aid us to understand the influence of the
differential rotation in the dynamics properties of the surrounding
space--time.

The potentials (\ref{Potenciales eje}) can be written in an alternative way, we mean
\begin{equation*}
e(z)=1+\displaystyle \sum_{i=3}^{3} \frac{e_{i}}{z-\beta_{i}}\,
,\qquad f(z)=\displaystyle \sum_{i=3}^{3}
\frac{f_{i}}{z-\beta_{i}}\, ,
\end{equation*}
with
\begin{eqnarray*}
e_{j}= (-1)^{j}\frac{2 m
\beta^{2}_{j}}{(\beta_{j}-\beta_{k})(\beta_{j}-\beta_{i})}\, ,\\
f_{j}= (-1)^{j+1}\frac{i \mu \beta_{j} + q\beta^{2}_{j}
}{(\beta_{j}-\beta_{k})(\beta_{j}-\beta_{i})}\, , \quad i,k \neq
j\, .
\end{eqnarray*}

Then, using (\ref{Ernst}) and (\ref{Potenciales eje}), following a
similar procedure to the one used in \cite{Breton},  we obtain the
Ernst potentials and the metric functions in whole space--time:
\begin{equation}\label{potenciales_ernst}
{\cal E}=\frac{A + B }{A - B}\, , \qquad \Phi=\frac{C}{A - B}\, ,
\end{equation}
\begin{eqnarray*}
f&=&\frac{A \bar{A}-B \bar{B} + C \bar{C}}{( A -
B)(\bar{A}-\bar{B})}\, ,\quad e^{2\gamma}=\frac{A \bar{A} -B
\bar{B} + C \bar{C}}{\displaystyle{K \bar{K}\prod_{n=1}^{6}r_n}}\, ,\\
\omega &=& \frac{{\rm Im}[(A + B)\bar{H}-(\bar{A} + \bar{B})G - C
\bar{I}]}{A \bar{A} - B \bar{B} + C \bar{C}}\, ,
\end{eqnarray*}
where
\begin{eqnarray*}
A &=& \displaystyle\sum_{{\small 1\leq i < j < k \leq 6}} a_{i j\,
k} r_{i}\,r_{j}\,r_{k}\, ,\quad B =\displaystyle\sum_{{\small
1\leq i < j \leq 6}} b_{i j} r_{i}\,r_{j},\\
C &=& \displaystyle\sum_{{\small 1\leq i < j \leq 6}} c_{i j}
r_{i}\,r_{j}\, ,\quad H = z\,A - (\beta_{1} + \beta_{2}+
\beta_{3})B \\
&+&\displaystyle \sum_{{\small 1\leq i < j < k \leq 6}} h_{i j\, k} r_{i}\,r_{j}\,r_{k} +
\displaystyle\sum_{{\small 1\leq i < j \leq 6}} (\alpha_{i} + \alpha_{j})\,b_{i j}
\,r_{i}\,r_{j},\\
G &=& -(\beta_{1} + \beta_{2} + \beta_{3})\,A + z\,B + \displaystyle\sum_{{\small 1\leq i < j \leq
6}} g_{i j} \,r_{i}\,r_{j}\\&+& \displaystyle \sum_{{\small 1\leq i < j < k \leq 6}} (\alpha_{i} +
\alpha_{j} + \alpha_{k})a_{i j\,k} r_{i}\,r_{j}\,r_{k},\\
I &=& (f_{1} + f_{2} + f_{3})(A - B) +
(\beta_{1}+\beta_{2} +
\beta_{3} - z)\,C \\
&+& \displaystyle\sum_{{\small 1\leq i < j < k \leq 6}} p_{i j\,k} r_{i}\,r_{j}\,r_{k} +
\displaystyle\sum_{{\small i=1}}^{6} p_{i}\, r_{i}\\ &+& \displaystyle\sum_{{\small 1\leq i < j
\leq 6}} [p_{i j}-(\alpha_{i} + \alpha_{j})c_{ij}] r_{i}\,r_{j},
\end{eqnarray*}
with
%%%%%%%%%%%%%%%%%%%%%%%%%%%%%%%%%%
%%Auxiliar definitions%%%%%%%%%%%%
\begin{eqnarray*}
\hspace{-1cm}
r_i &=& \sqrt{\rho^2 + (z-\alpha_i)^2}\, , \\
a_{i j\,k} &=& (-1)^{i + j + 1}\Lambda_{i j k}\,\Gamma_{l | m n}\, ,\\
b_{i j} &=& (-1)^{i + j}\lambda_{i j}\,H_{l | m n p}\, ,\\
c_{i j} &=& (-1)^{i + j}\lambda_{i j}[f(\alpha_l)\,\Gamma_{m | n
p} - f(\alpha_m)\,\Gamma_{n | p l}\\ &+& f(\alpha_n)\,\Gamma_{p |
l m}
- f(\alpha_p)\,\Gamma_{l | m n}]\, ,\\
h_{i j\,k} &=& (-1)^{i + j + k}\Lambda_{i j k}(e^{*}_1\,\delta_{2
3 | l m n} + e^{*}_2\,\delta_{3 1 | l m n}\\ &+&
e^{*}_3\,\delta_{1 2
| l m n})\, ,\\
g_{i j} &=& (-1)^{i + j}\lambda_{i j}(\alpha_{l}\,\Gamma_{m | n p}
- \alpha_{m}\,\Gamma_{n | p l}\\&+& \alpha_{n}\,\Gamma_{p | l m} -
\alpha_{p}\,\Gamma_{l | m n})\, ,\\
p_i &=& (-1)^i D_{i}[f(\alpha_l)\,H_{m | n p s} -
f(\alpha_m)\,H_{n | p s l}\\ &+& f(\alpha_n)\,H_{p | s l m} -
f(\alpha_p)\,H_{s | l m
n}\\ &+& f(\alpha_s)\,H_{l | m n p}]\, ,\\
p_{i j} &=& (-1)^{i + j} \lambda_{i j}(e^{*}_1\,\Upsilon_{2 3 | l
m n p} + e^{*}_2\,\Upsilon_{3 1 | l m n p}\\ &+&
e^{*}_3\,\Upsilon_{1
2 | l m n p})\, ,\\
p_{i j\,k} &=& (-1)^{i + j +1} \Lambda_{i j\,k}(e^{*}_1\,\Psi_{2 3
| l m n} + e^{*}_2\,\Psi_{3 1 | l m n}\\ &+& e^{*}_3\,\Psi_{1 2 |
l m n})\, ,
\end{eqnarray*}
and
\begin{eqnarray*}
\lambda_{i j} &=& (\alpha_i - \alpha_j)\,D_i\,D_j\, ,\\
\Lambda_{i j\,k} &=& (\alpha_i - \alpha_j)(\alpha_i -
\alpha_k)(\alpha_j - \alpha_k)\,D_i\,D_j\,D_k\, ,\\
D_i &=& \frac{1}{(\alpha_i - \beta_1)(\alpha_i - \beta_2)(\alpha_i
-
\beta_3)}\, ,\\
\Gamma_{l | m n} &=& H_3(\alpha_l)\,\Delta_{1 2 | m n} +
H_3(\alpha_m)\,\Delta_{1 2 | n l}\\ &+& H_3(\alpha_n)\,\Delta_{1 2
| l m}\, ,\\
\Delta_{l m | n p} &=& H_{l}(\alpha_n)\,H_{m}(\alpha_p) -
H_{l}(\alpha_p)\,H_{m}(\alpha_n)\, ,\\
H_{l}(\alpha_n) &=& \frac{2 \prod_{p \neq n} (\alpha_p -
\beta^{*}_l)}{\prod_{k \neq l}^{3} (\beta^{*}_l -
\beta^{*}_k)\,\prod_{k = 1}^{3} (\beta^{*}_l - \beta_k)}\\
&-& 2\,\sum_{k = 1}^{3} \frac{f^{*}_l\,f_k}{(\beta^{*}_l -
\beta_k)(\alpha_n - \beta_k)}\, ,\\
\delta_{l m | n p s} &=& \Delta_{l m | n p} + \Delta_{l m | p s} + \Delta_{l m | s n},\\
h_{l | m n p} &=& H_3(\alpha_l)\,\delta_{1 2 | m n p}\, ,\\
H_{l | m n p} &=& h_{l | m n p} + h_{m | n p l} + h_{n | p l m} +
h_{p | l m n}\, ,\\
\Psi_{l m | n p s} &=& f(\alpha_n)\,\Delta_{l m | p s} +
f(\alpha_p)\,\Delta_{l m | s n} + f(\alpha_s)\,\Delta_{l m | n p}\, ,\\
\Upsilon_{l m | n p r s} &=& f(\alpha_n)\,\delta_{l m | p r s} -
f(\alpha_p)\,\delta_{l m | r s n}\\&+&f(\alpha_r)\,\delta_{l m | s
n p} - f(\alpha_s)\,\delta_{l m | n p r}\, ,
\end{eqnarray*}
being $\alpha$'s the roots of the Sibgatullin's equation \cite{MetodoSibgatullin}
\begin{equation}\label{eq sibga}
e(z)+\tilde{e}(z)+2{\tilde{f}}(z)f(z)=0.
\end{equation}

%%%%%%%%%%%%%%%%%%%%%%%%%%%%%%%%%%%%
\section{Properties of the Solution}
\label{sec:Properties}%%%%%%%%%%%%%%
%%%%%%%%%%%%%%%%%%%%%%%%%%%%%%%%%%%%

%%%%%%%%%%%%%%%%%%%%%%%%%%%%%%%%%%%%%
\subsection{Multipolar structure}
\label{subsec:multipoles}
%%%%%%%%%%%%%%%%%%%%%%%%%%%%%%%%%%%%%

This solution describes the exterior gravitational field of a reflectionally symmetric source. Their
first four relativistic mass multipolar moments and first two electromagnetic multipolar moments
are arbitrary. The multipolar moments are calculated using the Fodor-Hoenselaers-Perj\'es procedure
\cite{HoensPerj} following the parametrization given in (\ref{Potenciales eje}), which has the
advantange that the real parameters $m$ corresponds to the total mass $M$, $a$ to the total
angular moment per unit mass ($a=J/m$, being $J$ the total angular moment) and $q$ to the total electric
charge $Q$; while $k$, $s$ and $\mu$ represent the mass-quadrupole moment $M_2$, current octupole
$S_3$ and magnetic dipole $\cal B$ through the following formulas:
\begin{eqnarray}
M_2 &=& m k - m a^2, \nonumber\\
S_3  &=& 2 i m a k  - i a m^3 - i m s, \nonumber\\
{\cal B}&=& i \mu + i a q. \label{multipoles}
\end{eqnarray}
The solution also has electric charge quadrupole moment ${\cal Q}_2$, which is not independent because
it depends on the other multipoles, given by the formula
\begin{equation} \label{quadrupole charge} {\cal Q}_2 = - a
\mu - a^2 q  + k q\, ,
\end{equation}
and permits that the solution has electric field even when $q$ vanish (induced electric field by
rotation \cite{framedragging}).

For NSs the mass--quadrupole moment is, surprisingly, approximated by the simple quadratic
relation \cite{Laarakkers}
\begin{equation}
M_2\approx-c(M, EOS)\frac{J^2}{M},
\end{equation}
where the constant $c=c(M,EOS)$ depends only on the mass M and the equation of state (EOS) for the
interior of the NS.  So, the formulas (\ref{Potenciales eje}) are successfull to fit $M_2$ of the
exact solucion with data of numerical solutions because of the form of $M_2$ and
the freedom to set the numerical value of the real parameter $k$. In addition, the absence of the
gravitomagnetic parameter (monopole of angular momentum) leads to achieve the asymptotic flatness
condition of the solution, necessary to describe a compact object.

%%%%%%%%%%%%%%%%%%%%%%%%%%%%%%%%%%%%%
\subsection{Limiting Cases}
%%%%%%%%%%%%%%%%%%%%%%%%%%%%%%%%%%%%%

A special feature of this solution is that in rigid rotation, i.e., with $s=0$, it contains the
following well-known limiting cases:
\begin{enumerate}
\item The stationary vacuum case $q=0$, $\mu=0$ with non-vanishing arbitrary deformation parameter
$k=-1/4 (m^2 - a^2)$ represents the Tomimatsu-Sato $\delta=2$ solution \cite{TSD2} with the mass--
quadrupole moment $M_2=-1/4 (m^3 - J^2/m)$.
%%%%%%%%%%%%%%%%%%%%%%%%%%%%%%%%%%%%%
\item The magnetostatic limit $q=0$, $a=0$ represents the massive magnetic dipole solution of Bonnor
\cite{Bonnor}.
%%%%%%%%%%%%%%%%%%%%%%%%%%%%%%%%%%%%%
\item The Kerr-Newmann space--time is obtained with $\mu=k=0$.
%%%%%%%%%%%%%%%%%%%%%%%%%%%%%%%%%%%%%
\item The stationary vacuum case $q=0$, $\mu=0$ with non-arbitrary
deformation, i.e., with $k=0$ represents the Kerr space--time.
\end{enumerate}
It is important to emphasize that the present six parametric
solution generalizes the Kerr and Kerr--Newmann solutions,
which are the most studied stationary axisymmetric solutions.

%%%%%%%%%%%%%%%%%%%%%%%%%%%%%%%%%%%%%%%
\subsection{Electromagnetic Fields}
%%%%%%%%%%%%%%%%%%%%%%%%%%%%%%%%%%%%%%%
\suppressfloats

The strong magnetic field is one of the most important features of NSs and pulsars. The last ones have
non-aligned dipole moment with the rotation axis causing loss of energy and angular
moment, so the star radiates electromagnetic and gravitational waves away and decrease its velocity
of rotation, configuring a clear example of a non­stationary and non­axisymmetric system. On
the other hand, the stationary axisymmetric case is characterized by the fact that the magnetic
dipole moment is aligned with the axis rotation of the star and the magnetic field is poloidal
(i.e. the magnetic field lying in the meridional planes \cite{BonazzolaBocquet}). Our exact
solution is stationary and axisymmetric and, therefore, describes this kind of NSs.

For the description of the electromagnetic properties of the solution, we must to calculate the
electric and magnetic fields produced in the surrounding space--time. The reason to introduce
electric field in the solution, in spite of the fact of global neutrality of the astrophysical
objects, is that an electric field is induced by the rotation of the magnetized objects, as Eq.
(\ref{quadrupole charge}) shows. To have  a closed form the electric and magnetic fields is important, for example, in order to
study the accretion of charged particles around of the NS,  to study geodesic motion of test
particles around a NS and so on. The electric and magnetic fields can be calculated using the
expressions
\begin{equation}\label{EB}
E_{\alpha} = F_{\alpha \beta}u^{\beta}\, ,\qquad B_{\alpha} = -\frac{1}{2}\,\epsilon_{\alpha \beta}^{\hphantom{\alpha \beta}\gamma \delta} F_{\gamma \delta}u^{\beta}\, ,
\end{equation}
where $F_{\alpha \beta}$ is the electromagnetic field tensor $F_{\alpha \beta} =
2\,A_{[\beta;\alpha]}$, $A_{\mu}=(0,0,A_{\phi},-A_t)$ is the electromagnetic four-potential,
$\epsilon_{\alpha \beta \gamma \delta}$ is the totally antisymmetric tensor the positive orientation with norm $\epsilon_{\alpha \beta \gamma \delta}\epsilon^{\alpha \beta \gamma \delta} = -24$ \cite{wald} and $u_{\alpha}$ is a time--like vector (in
the case of fluid, this vector is the four--velocity of the fluid). For a congruence of observers at rest in the frame of (\ref{Papapetrou}), the four--velocity is defined by the time--like vector
\begin{equation}
u^\alpha=(1/\sqrt{f},0,0,0),
\end{equation}
so the vectorial fields have components in the $\rho$ and $\phi$ directions.  The electric field is given by the following expressions:
\begin{equation}\label{Eexplicito}
E_\rho = -\frac{\sqrt{f}}{e^{2 \gamma}}\,A_{t, \rho}\, ,\qquad E_z
= -\frac{\sqrt{f}}{e^{2 \gamma}}\,A_{t, z}\, ,
\end{equation}
and the magnetic field by
\begin{eqnarray}\label{Bexplicito}
B_\rho &=& \frac{f^{3/2}}{\rho e^{2 \gamma}}\,(-\omega A_{t,z} +
A_{\phi,z})\, ,\\B_z &=& -\frac{f^{3/2}}{\rho e^{2
\gamma}}\,(-\omega A_{t,\rho} + A_{\phi,\rho})\, .
\end{eqnarray}
The $A_t$ potential is the real part of the electromagnetic Ernst
potential $\Phi$ given by (\ref{potenciales_ernst}), and the
potential $A_{\phi}$ can be calculated as the real part of the
Kinnersley potential ${\cal K}= A_{\phi}+iA'_{t}$ \cite{kinn}, which
can be obtained using the Sibgatullin method and can be written as
\begin{equation}\label{kinnersleypotential} {\cal
K}=-i\,\frac{I\,(f_{1} + f_{2})}{A - B}\, .
\end{equation}
Thus, the closed--form expressions for the electric and magnetic fields are
\begin{widetext}
\begin{subequations}\label{E}
\begin{eqnarray}
E_\rho &=& -\frac{|K|^2\prod_{n=1}^{6}r_n}{|A - B|\sqrt{|A|^2 -
|B|^2 + |C|^2}}\,{\rm Re}\Bigg\{ \left[ \frac{ C_{,\rho} -
C\ln (A - B)_{,\rho}}{A - B} \right] \Bigg\}\, ,\\
E_z &=& -\frac{|K|^2\prod_{n=1}^{6}r_n}{|A - B|\sqrt{|A|^2 - |B|^2 +
|C|^2}}\,{\rm Re}\Bigg\{ \left[ \frac{ C_{,z} - C\ln (A - B)_{,z}}{A
- B} \right] \Bigg\}\, ,
\end{eqnarray}
\end{subequations}
\begin{subequations}\label{B}
\begin{eqnarray}
&B_\rho& = \frac{{\rm Im} (A - B)\bar{H} +(\bar{A} - \bar{B})G -
C\bar{I}]}{\rho |A - B|^2}E_z+\frac{|K|^2\,\sqrt{|A|^2 - |B|^2 +
|C|^2}\,\prod_{n=1}^{6}r_n}{\rho |A - B|^3}\,{\rm
Im}\Bigg\{\frac{(\bar{f_1} + \bar{f_2})\left[ \bar{I}_{,z}
-\bar{I}\ln (\bar{A} - \bar{B})_{,z}\right] }{A-B} \Bigg\}
\nonumber\ ,\\&& \\&B_z& = - \frac{{\rm Im} (A - B)\bar{H} +(\bar{A}
- \bar{B})G - C\bar{I}]}{\rho |A -
B|^2}E_{\rho}+\frac{|K|^2\,\sqrt{|A|^2 - |B|^2 +
|C|^2}\,\prod_{n=1}^{6}r_n}{\rho |A - B|^3}\,{\rm
Im}\Bigg\{\frac{(\bar{f_1} + \bar{f_2})\left[ \bar{I}_{,\rho}
-\bar{I}\ln (\bar{A} - \bar{B})_{,\rho}\right] }{A-B}
\Bigg\}.\nonumber\\&&
\end{eqnarray}
\end{subequations}
\end{widetext}
In order to show that our solution have the appropriate features of realistic NSs, we have to use
typical values of NS for their physical parameters, like mass, rotation, deformation, charge and
magnetic dipole moment. NS are compact objects that possess strong magnetic fields ($B \sim 10^{12}
T$), nevertheless, in our solution the parameter related  with the magnetic field is the magnetic
dipole moment $\mu$.  In \cite{SlowlyJD} Manko {\it et. al.} showed how the magnetic dipole moment
of their exterior solution depends on the values of the parameters of the interior solution, such
as the matter density, the current function and the stellar radius.  They obtained the above result
by comparison between their magnetic potential $A_{\phi}$ at infinite and the Ferraro's potential
\cite{ferraro}; this became a test for the weak field because the Ferraro's solution is an approximate
one that have into account deformation only at first order around the spherical symmetry
\cite{BonazzolaBocquet}. According to this, we must to use typical values given for realistic
interior solutions.  Bocquet \emph{et al.}
\cite{BonazzolaBocquet} calculated interior solutions
for the Einstein--Maxwell system using the typical value of $10^{32} A\,m^2$ for the magnetic
dipolar moment of the NS in S.I. units. In natural units ($c=G=\mu_0=\varepsilon_0=1$), which have
been used in our paper, the magnetic dipole moment has a order of
\begin{equation}\label{conversion}
\mu_{natural} \sim \frac{10^{-6}\sqrt{\mu_0 G}}{c^2}\,\mu_{S.I.}\,,
\end{equation}
resulting a typical value for the $\mu$ parameter of our solution of order $10^1 Km^2$.

\begin{center}
\begin{figure}[h]
\begin{center}
\begin{tabular}{c}
(a)\\
\hspace{-0.5cm}
\parbox{7cm}{\includegraphics[width=8.8cm,height=8.5cm]{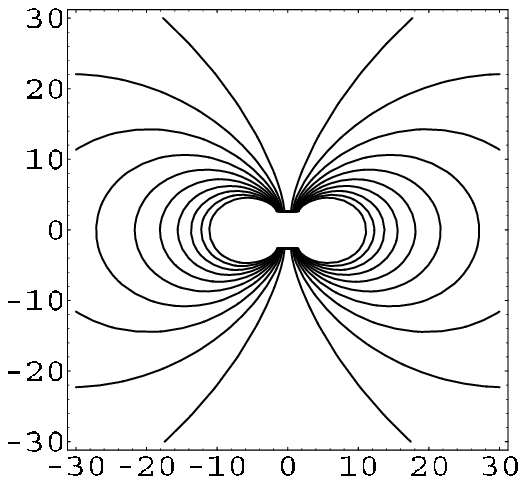}}\\
(b)\\
\parbox{7cm}{\includegraphics[width=8.8cm,height=8.4cm]
{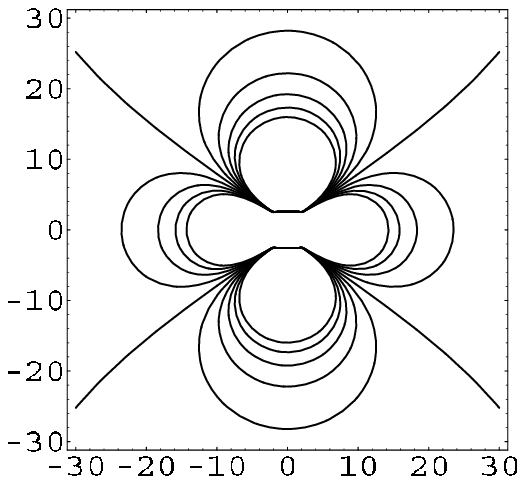}}
\end{tabular}
\end{center}
\caption{(a) Magnetic field force lines and (b) isopotential lines
of the induced electric field for $m =1.742M_{\odot}$,  $a = 1.009
Km$, $k = -0.336 Km^2$, $s = 0.212 Km^3$, $q = 0 Km$,  $b = 10
Km^2$.} \label{Fig:lineasdecampo}
\end{figure}
\end{center}

In the Fig. \ref{Fig:lineasdecampo} we have plotted the force lines  of the magnetic field and
the isopotential lines of the induced electric field ($A_t=$constant) for one possible model of NS
($M_{\odot} = 1.47663 Km$). The increasing of the separation between consecutive force lines
indicates, as is usual, that the electric field and the magnetic one decreases while the distance
increases. As we expected, in absence of monopole electric charge distribution, the electric field is
quadrupolar. In addition, the same figure shows that the axistationary properties of the magnetic
field appear: aligned magnetic dipole moment and dipolar structure (poloidal field).
%%%%%%%%%%%%%%%%%%%%%%%%%%
\subsection{Singularities}
\begin{figure}[h]
\begin{center}
\begin{tabular}{c}
\hspace{-2.5cm}
\parbox{7cm}{\includegraphics[width=8.8cm,height=8.8cm]{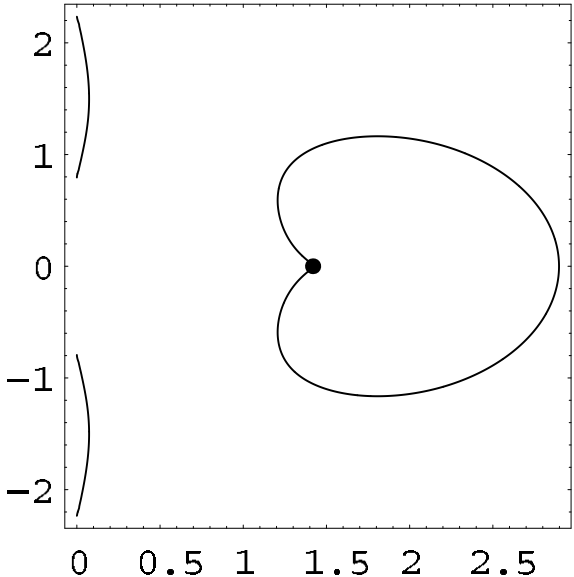}}
\end{tabular}
\end{center}
\caption{Stationary limit surfaces for $m
=2.388M_{\odot}=3.526 Km$, $a = 2.264 Km$, $k = -1.808 Km^2$, $s =
2.498 Km^3$.} \label{singularidades}
\end{figure}

%%%%%%%%%%%%%%%%%%%%%%%%%%
The singularities are solutions of the equation
\begin{equation}
A - B =0\, ,
\end{equation}
i.e., denominator of the Ernst potentials given by
(\ref{potenciales_ernst}) equals to zero. In the Fig. \ref{singularidades} we
have plotted the typical shapes of stationary limit surfaces ($f=0$) for specific sets of
realistic parameters given in Ref. \cite{BertiStergioulas}. The location of ring
singularity appears with dots. The ring singularity lie on the stationary
limit surface, which is a characteristic of the stationary vacuum solutions.

For all stellar
models there are no ring singularities outside of infinite red shift surfaces, therefore the
solution can represent the exterior field of compact objects. In addition, we
show in this section, with an approximate calculation via Komar masses, that the ring
singularities of solution are massless.
The geometrical structure of the exact solution implies that one could consider it as superposition
of subextreme and hyperextreme objects.  The Komar mass $M_i$ of a subextreme part can be calculated
using the Tomimatsu formula \cite{MasaTomimatsu}
\begin{equation}\label{subextreme} M_i =
\frac{1}{4} \int_{\alpha_<}^{\alpha_>} \omega ^{(i)} \Omega_{,z} dz\, ,
\end{equation}
where $\Omega$ is the imaginary part of the Ernst potential
$\varepsilon$, and $\omega ^{i}$ is the constant value of the
metric coefficient $\omega$ on the segment $[\alpha_<,\alpha_>]$
of the $z$--axis.

In the case of hyperextreme part, the Komar mass cannot be
evaluated using the formula (\ref{subextreme}), and one more general integral expression should be
used \cite{MMRuiz}
\begin{eqnarray}\label{hyperextreme}
M_i &=&\frac{1}{4} \left\{\int\limits _{z_l}^{z_u} [\rho (\ln
f),_\rho - \omega \Omega,_z]_{\rho = \rho_0} dz + \int\limits
_{0}^{\rho_0} [\rho (\ln f),_z \right.\nonumber\\
&-\omega& \left. \Omega,_\rho]_{z = z_u} d\rho -\int\limits
_{0}^{\rho_0} [\rho (\ln f),_z - \omega \Omega,_\rho]_{z = z_l}
d\rho\right\}\, ,
\end{eqnarray}
where the integration is carried out over the surface of a cylinder enclosing the cut which joins
the points $\alpha_i$ and $\bar{\alpha}_i$; $z_u$ and $z_l$ denote locations on the symmetry axis
of the centers of upper and lower bases, $\rho_0$ is the radius of the bases.  We
will illustrate the procedure followed using the set of parameters showed in the Fig.
\ref{singularidades}. First, we shall calculate the mass of the NS via Komar masses and then we
will compare the obtained result with the mass monopole $m$, to deduce that the mass of the ring
singularity is equal to zero.  For the set of parameters showed in Fig.~\ref{singularidades}, the
roots of the Sibgatullin equation (\ref{eq sibga}) are $\alpha_1=0.792, \alpha_2=2.245,
\alpha_3=-0.792, \alpha_4=-2.245, \alpha_5=-1.406 i, \alpha_6=\bar{\alpha}_5$, producing four
subextreme parts and one hyperextreme part. Using the formula (\ref{subextreme}), the mass of the
subextreme contribution is $m_{s}=1.523 Km$ and using the formula (\ref{hyperextreme}), the mass of
the hyperextreme contribution is $m_{h}=2.003 Km$, then $m_{s}+m_{h}=3.526 Km=m$.   We made the
above calculations for all data for all EOS used from Ref. \cite{BertiStergioulas} and the results
were allways the same.

As can be seen, the calculations were made in a numerical way for
the different stellar modells;  it brings us to give only
approximate values for the mass of ring singularity. For that
reason, the ring singularity arise with a mass equals to zero with
precision up to $10^{-4}$, with no dependence of the equation of
state used.

%%%%%%%%%%%%%%%%%%%%%%%%%%%%%%%%%%%%%%%%%%%%%%%%%%%%%%%%%%%%%%
\section{Test of accuracy of the six--parameter analytic model}
\label{sec:checkSol}%%%%%%%%%%%%%%%%%%%%%%%%%%%%%%%%%%%%%%%%%%

In order to demonstrate that our solution can represent the exterior gravitational field of a
realistic NS, we have to do comparisons with results obtained from numerical interior solutions and
with others possible analytic models. We have chosen the Kerr solution, the
expressions derived by Shibata \& Sasaki (S\&S) in Ref.\cite{Shibata} and Hartle-Thorne's solution
(H\&T) truncated to second order of the expansion parameter $\epsilon=\Omega/\Omega*$
(see Ref.\cite{Abramowicz} and \cite{BertiBruni}). For the comparison
with numerical models, we have chosen the solutions by B\&S Ref.\cite{BertiStergioulas}; the comparison
procedure was taken from there, that is, we compared an invariant quantity of the space--time, the multipolar moments, and other
quantity which can be depending on the coordinates. The depending coordinates quantity could be the
gravitational red shift or the radii of the Innermost Stable Circular Orbits (ISCOs). For this work
we have chosen to compare the radii of the ISCO because it is related to several astrophysical
properties of rapidly rotating NSs, e.g. the accretion disk cannot be longer that radii of the ISCO,
and this sets an upper limit to the Keplerian frequency of particles orbiting a star.

We matched the exact solution with the numerical interior solutions imposing the condition that the
first four mass and current multipole moments, i.e. mass, angular moment, mass--quadrupole and current octupole,
of the exact solution and the numerical ones have the
same value.  In B\&S, sources possessing electric charge or magnetic dipolar moment were not
considered, so in this work the comparison of the solutions was developed in the vacuum case.

The numerical solutions were calculated for five interior numerical solutions for the
gravitational field of a  NS, each numerical solution is described by a different Equation Of State
(EOS). The EOS used were denoted as A \cite{EOSA}, AU \cite{EOSAU}, FPS
\cite{EOSFPS}, L \cite{EOSL}, APRb \cite{EOSAPRb} (see B\&S \cite{BertiStergioulas} and Cook {\it et al.} \cite{Cook94} for details).
These numerical solutions were calculated for three different sequences of constant rest
mass:
\begin{itemize}
\item First sequence. The sequence corresponding to a canonical neutron star having gravitational
mass $M=1.4 M_\odot$ in the non rotating limit.
%%%%%%%%%%%%%%%%%%%%%%%%%%%%%%%%%%%%%
\item Second sequence. The sequence terminating at the maximum-mass model in the non rotating limit
(maximum-mass normal sequence).
%%%%%%%%%%%%%%%%%%%%%%%%%%%%%%%%%%%%%
\item Third sequence. A supramassive sequence, i.e., a sequence which does not terminate at a
non--rotating model.
\end{itemize}
In order to distinguish when a neutron star is rapidly rotating or not, we shall use the
dimensionless parameter $j$, defined by $j\equiv J/M^2,$ as it was used by Sibgatullin \& Sunyaev \cite{Sibgatullin},
Stute \& Camenzind \cite{StuteCamenzind} and  B\&S \cite{BertiStergioulas}. In spite of $j$ is not a direct
quantitative measure of the magnitude of the rotation rate of the stars, it gives us a measure of
the angular momentum which relates proportionally to the angular velocity.

To clarify the
meaning of the value of the $j$ parameter lets consider an example. For the recently discovered
pulsar \cite{716Hz} which frequency is 716--Hz, the value of $j$ depends on the EOS and the
sequence of mass considered. Using the values reported in B\&S for stars with similar frequency, we
find that $j$ value must be around $0.42<j<0.68$. For example, for the EOS FPS in the first
sequence of mass and assuming $m \sim 1.430M_\odot$ and $J \sim 2.961\, Km^2$, the value of $j$
must be very close to $0.67$, and for the second sequence assuming $m \sim 1.831M_\odot$ and $J
\sim 3.185\, Km^2$, the value of $j$ must be very close to $0.43$. Although, it is clear that we
can not present a reliable classification for rapidly rotating neutron stars based on $j$,
we shall assume that a NS is considered rapidly rotating when $j>0.35$.
%%%%%%%%%%%%%%%%%%%%%%%%%%%%%%%%%%%%%%%%%%%%%
\subsection{Innermost stable circular orbits}
%%%%%%%%%%%%%%%%%%%%%%%%%%%%%%%%%%%%%%%%%%%%%
It is well known that not all orbits around relativistic stars are stable. For non-rotating stars,
the ISCO is located at a circumferential radius of $R_{ISCO}=6M$ (see Ref.~\cite{bardeen} for a
complete treatment of circular orbits in stationary axisymmetric spacetimes).  Depending on the EOS and the mass
of the star, the ISCO can be located outside the stellar surface.  The rotation introduces a preferred
direction in the $\phi$ coordinate, so ISCOs around a rotating star belong to two families: the
co-rotating and the counter-rotating one.

A circular orbit in the equatorial plane is one for which $\rho
={\rm const.}$ The equation for geodesic motion along the radial coordinate $\rho$ reads
\begin{equation}
-g_{\rho \rho} \left(\frac{d\rho}{d\tau}\right)^2 = 1-\frac{E^2
g_{\phi\phi}+2E\,L g_{t\phi}+L^2 g_{tt}} {g^2_{t\phi} -
g_{tt}\,g_{\phi\phi}}\equiv V(\rho),
\end{equation}

where $E$ and $L$ are the conserved energy and angular momentum per unit mass, determined by the
conditions $V=dV/d\rho=0$. Geodesics become marginally stable when $d^2 V/d\rho^2=0$, so it is
\begin{eqnarray}
&\omega'&  \omega'' f^5 \rho(2f-f' \rho)+\omega'^2f^4 [2f^2+(-f'^2
+f'' f)\rho^2 ]\nonumber \\&+& \omega' f^2 \sqrt{\omega'^2f^4+f'
\rho(2f-f'
\rho)}[2f^2+2f'^2\rho^2\nonumber \\
&-& f\rho(4f'+f'' \rho) + \rho(2f-f' \rho) \bigl\{3f' f^2\nonumber \\
&-&4f'^2 f\rho+f'^3\rho^2+f^2[f'' \rho \nonumber \\
&-&\omega'' f\sqrt{\omega'^2f^4+f' \rho(2f-f' \rho)} ] \bigr\} =0,
\end{eqnarray}
for co--rotating orbits ($R_+$), where $'$ indicates a partial derivative with respect to $\rho$.
For the case of Hartle-Thorne's solution, the radii of the ISCO for co--rotating orbits ($R_+$) was
derived by Abramowicz {\it et al.} in \cite{Abramowicz}:
\begin{eqnarray}\label{RiscoHT}
R^{H\&T}_{ISCO}&=& 6m \left[1 -\sqrt{\frac{8}{27}}j+
\left(\frac{251647}{2592}-240 \ln{\frac{3}{2}}\right)j^2-
\nonumber \right.\\ &-& \left. \left(\frac{9325}{96}-240
\ln{\frac{3}{2}}\right)q \right]\, ,\\ \nonumber
\end{eqnarray}
where $q$ is a parameter related with the cuadripolar deformation and defined as $q=Q/m^3$, where $Q$
is the cudripolar deformation multipole.

S\&S derived a general representation of axisymmetric vacuum solutions (in the
form of a series expansion of the physical multipole moments of the space--time) and found some
approximate analytic formulas for the location of the inner stable circular orbit, the angular
moment, and energy of a particle around a relativistic source \cite{Shibata}. In general, their
formulas depend on the mass, angular momentum, mass-- quadrupole, current octupole and mass
$2^4$-pole moments and on upper order multipoles in the rotation parameter. Including all terms up
to order $O(4)$ in the rotation parameter, they found the following equation for the circumferential
radius of the corotating ISCO ($R_+$):
\begin{eqnarray}\label{SSISCO}
R^{S\&S}_{ISCO}&=&6m(1-0.54433j-0.22619j^2+0.17989q_2\nonumber\\
&-&0.23002j^3+0.26296jq_2-0.05317q_3\nonumber\\
&-&0.29693j^4+0.44546j^2q_2-0.06249q_2^2\nonumber\\
&+&0.01544q_4-0.11310jq_3).
\end{eqnarray}
In the previous expression, dimensionless parameters $q_2=-M_2/m^3$, $q_3=-S_3/m^4$ and $q_4=M_4/m^5$
have been introduced. S\&S adopted the approximation, $q_4=\alpha_4 q_2\,^2$ because
the numerical calculation of the $2^4$-pole term is very hard and difficult. The value of $\alpha_4$
lies between 0 and 2, we adopted $\alpha_4=1$ as it is generally used \cite{Shibata}. In the case of the Kerr metric,
the approximate expression for the location of the co--rotating
ISCO up to order $O(j^4)$ is
\begin{eqnarray}\label{kISCO}
R^{Kerr}_{ISCO}&=&6M(1-0.54433j-0.04630j^2\\
&-&0.02016^3-0.01110j^4).\nonumber
\end{eqnarray}
In all cases, the location of the counter--rotating ISCO ($R_-$) is obtained simply using the above
equation and changing the sign of all the star's rotation multipoles. The obtained results from the direct
comparison between the ISCOs are presented in the
Fig. \ref{Fig:AjusteROME1}.

\vspace{-4.5cm}
\begin{widetext}
\begin{center}
\begin{figure}[!hbt]
\begin{tabular}{cc}
\vspace{-0.5cm} (a.1) & (a.2)\\ \vspace{-0.5cm}
\hspace{-0.7cm}\includegraphics[width=9.3cm,height=8.4cm]{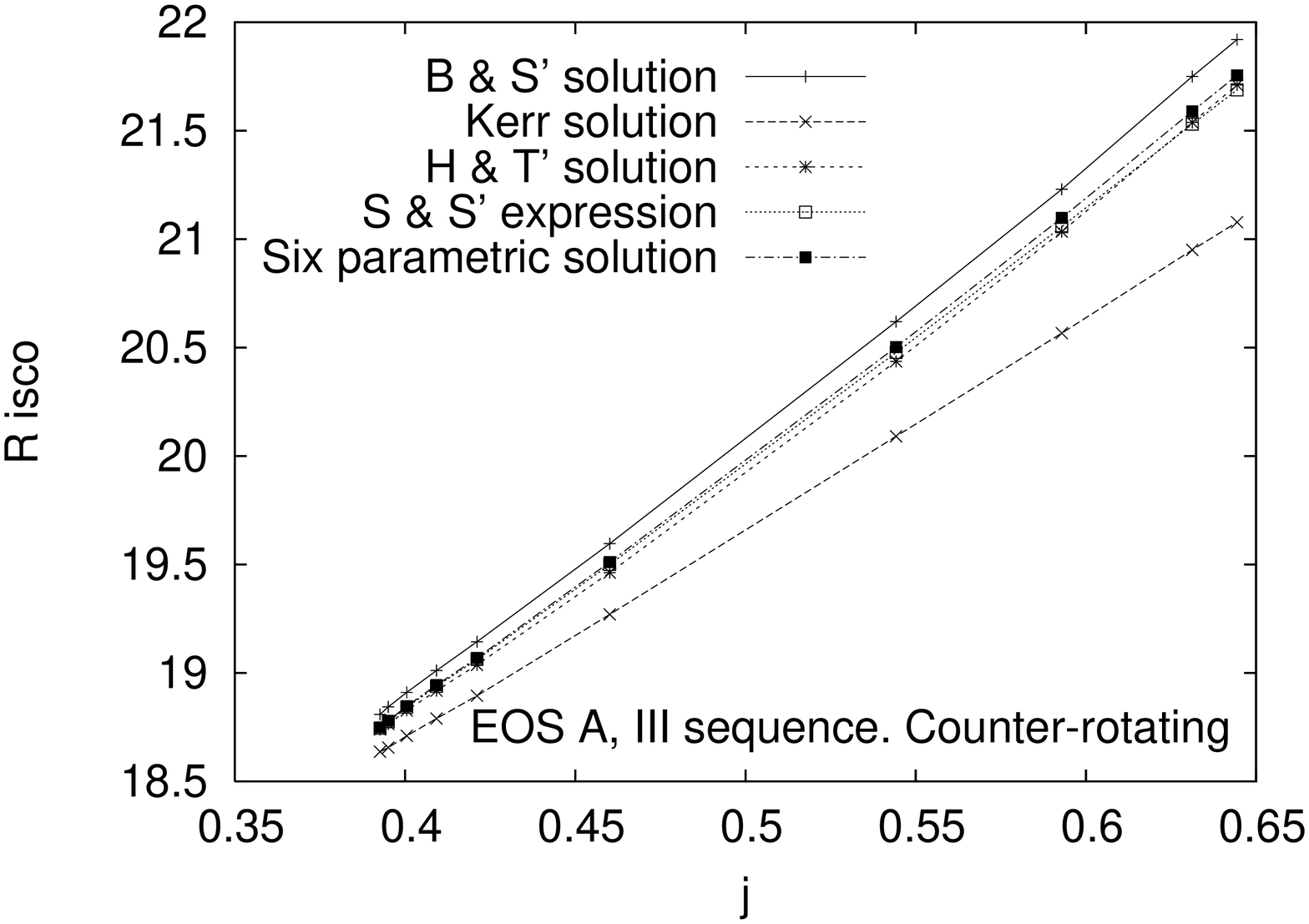}&
\hspace{-0.7cm}\includegraphics[width=9.3cm,height=8.4cm]{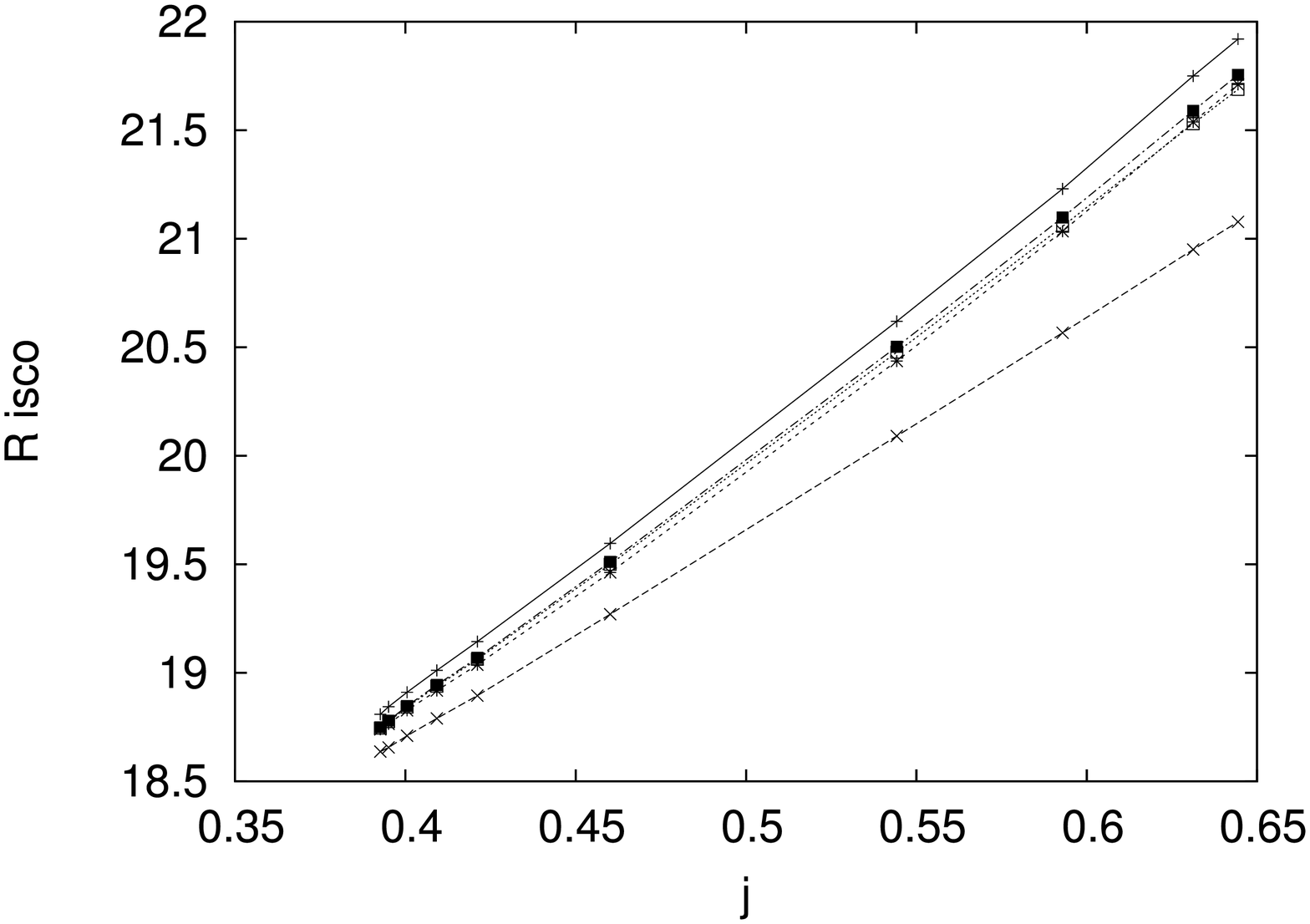}\\
\vspace{-0.5cm} (b.1) & (b.2)\\ \vspace{-0.5cm}
\hspace{-0.7cm}\includegraphics[width=9.3cm,height=8.4cm]{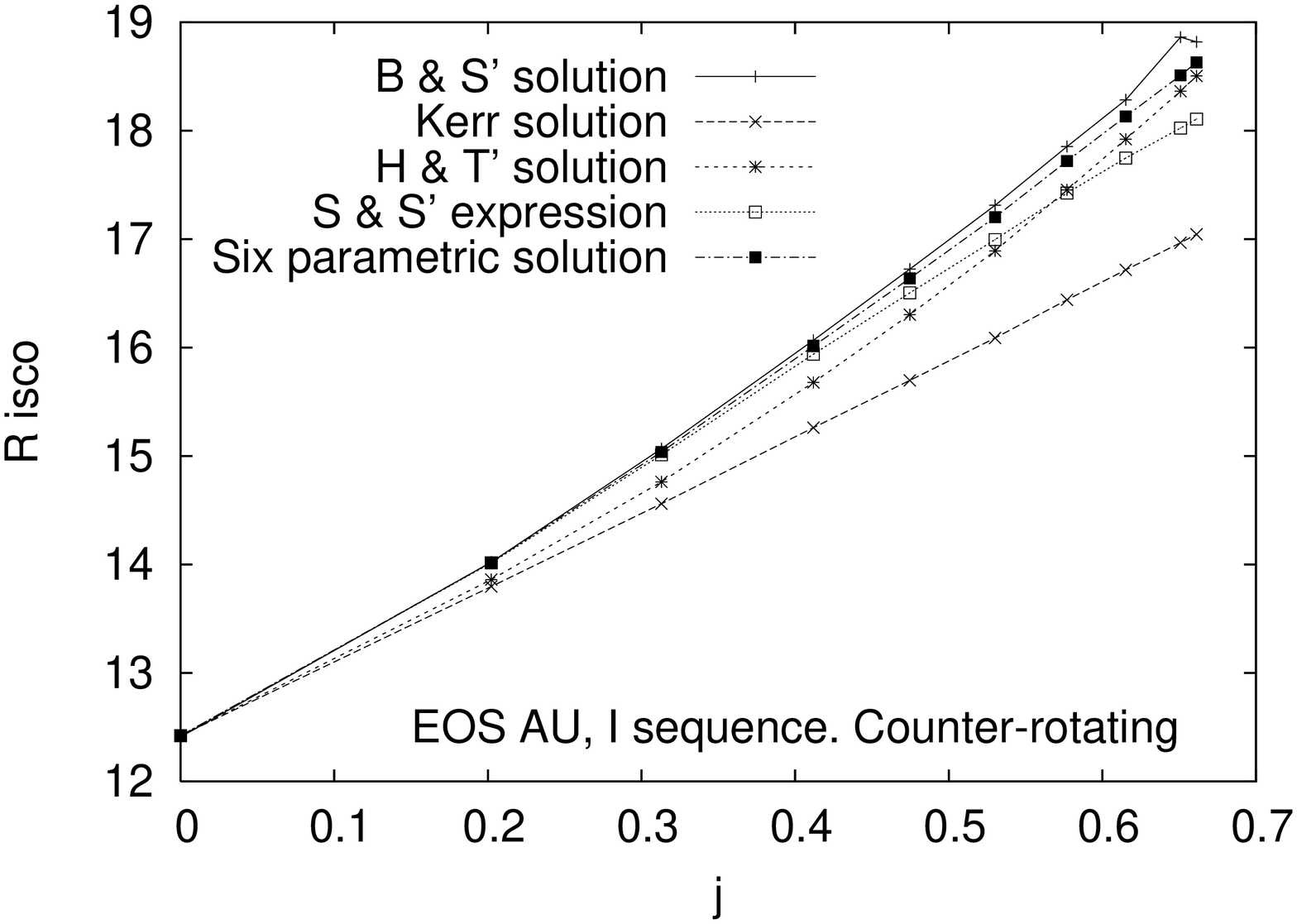}&
\hspace{-0.7cm}\includegraphics[width=9.3cm,height=8.4cm]{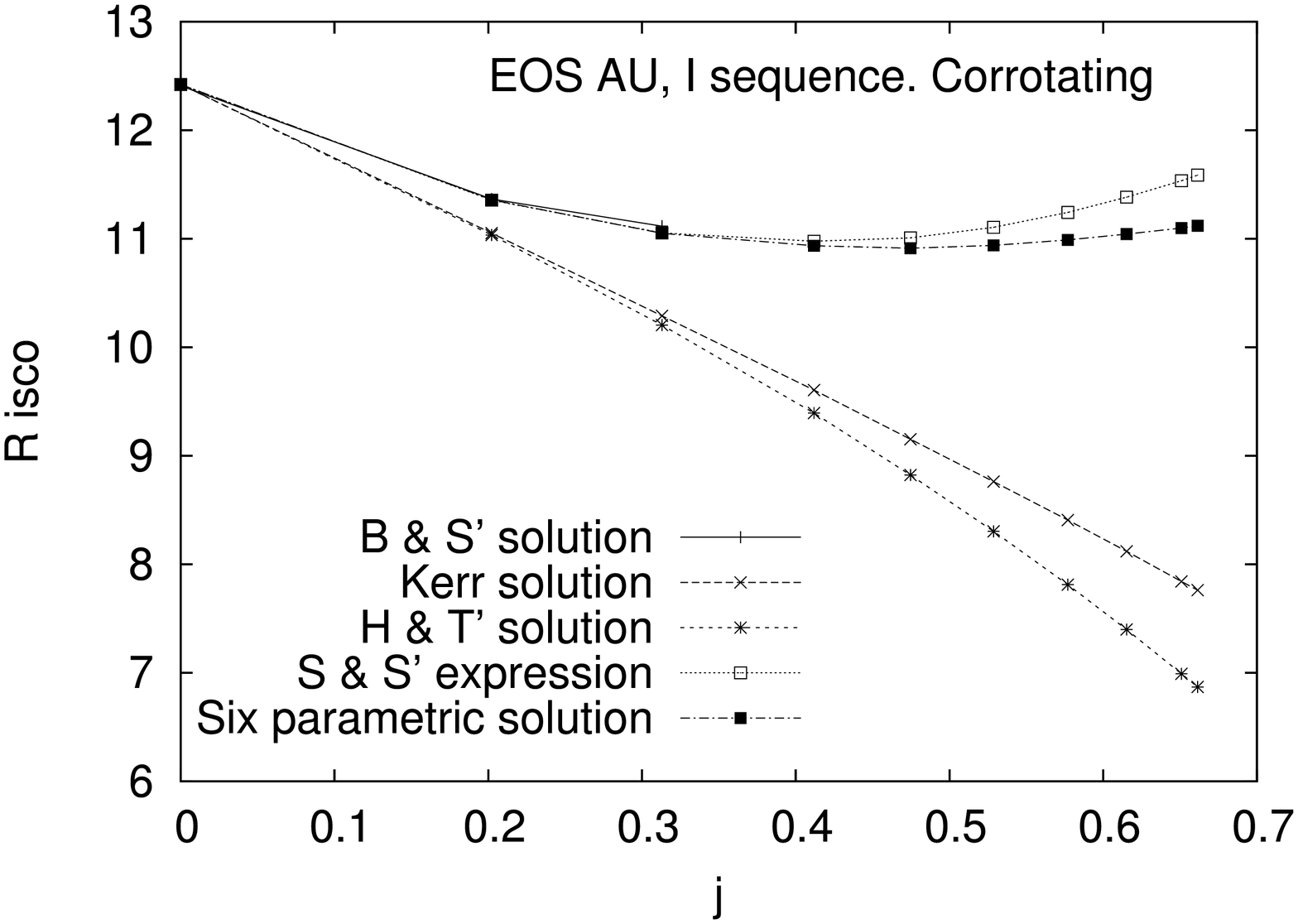}\\
\vspace{-0.5cm} (c.1) & (c.2)\\
\hspace{-0.7cm}\includegraphics[width=9.3cm,height=8.4cm]{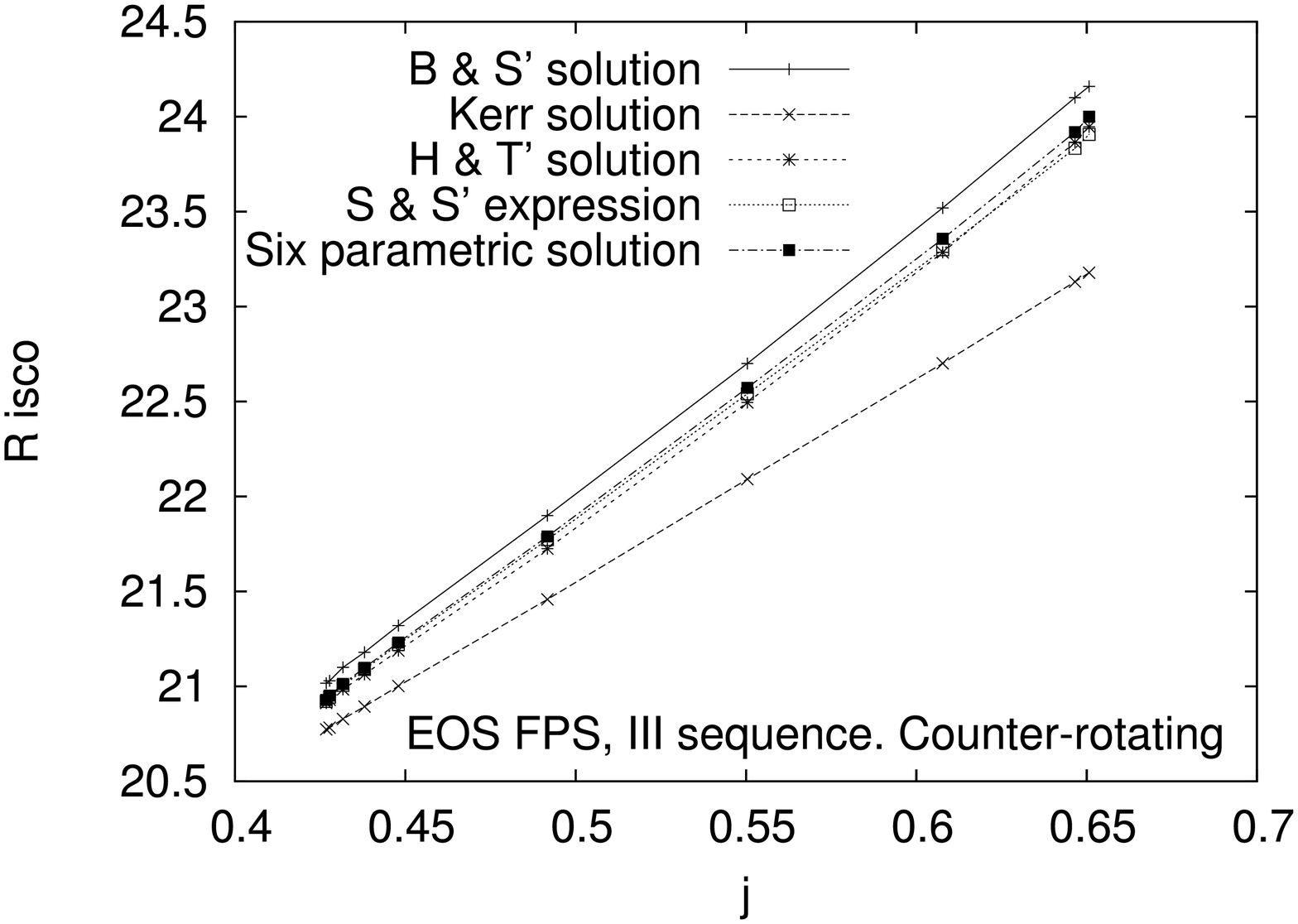}&
\hspace{-0.7cm}\includegraphics[width=9.3cm,height=8.4cm]{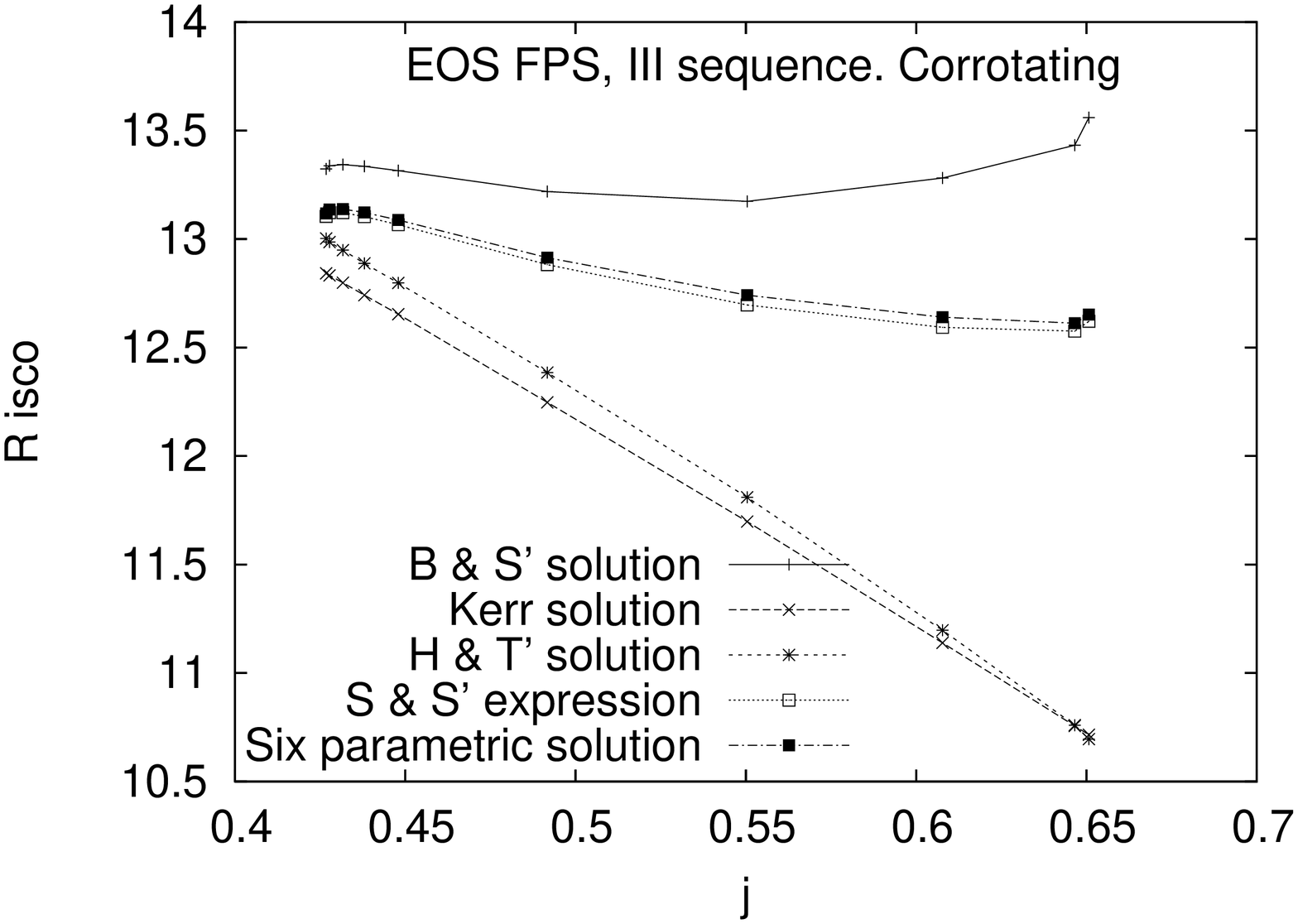}
\end{tabular}
\end{figure}
\end{center}
\clearpage
\begin{center}
\begin{figure}[!hbt]
\begin{tabular}{cc}
\vspace{-0.5cm} (d.1) & (d.2)\\
\hspace{-0.5cm}\includegraphics[width=8.8cm,height=8.4cm]{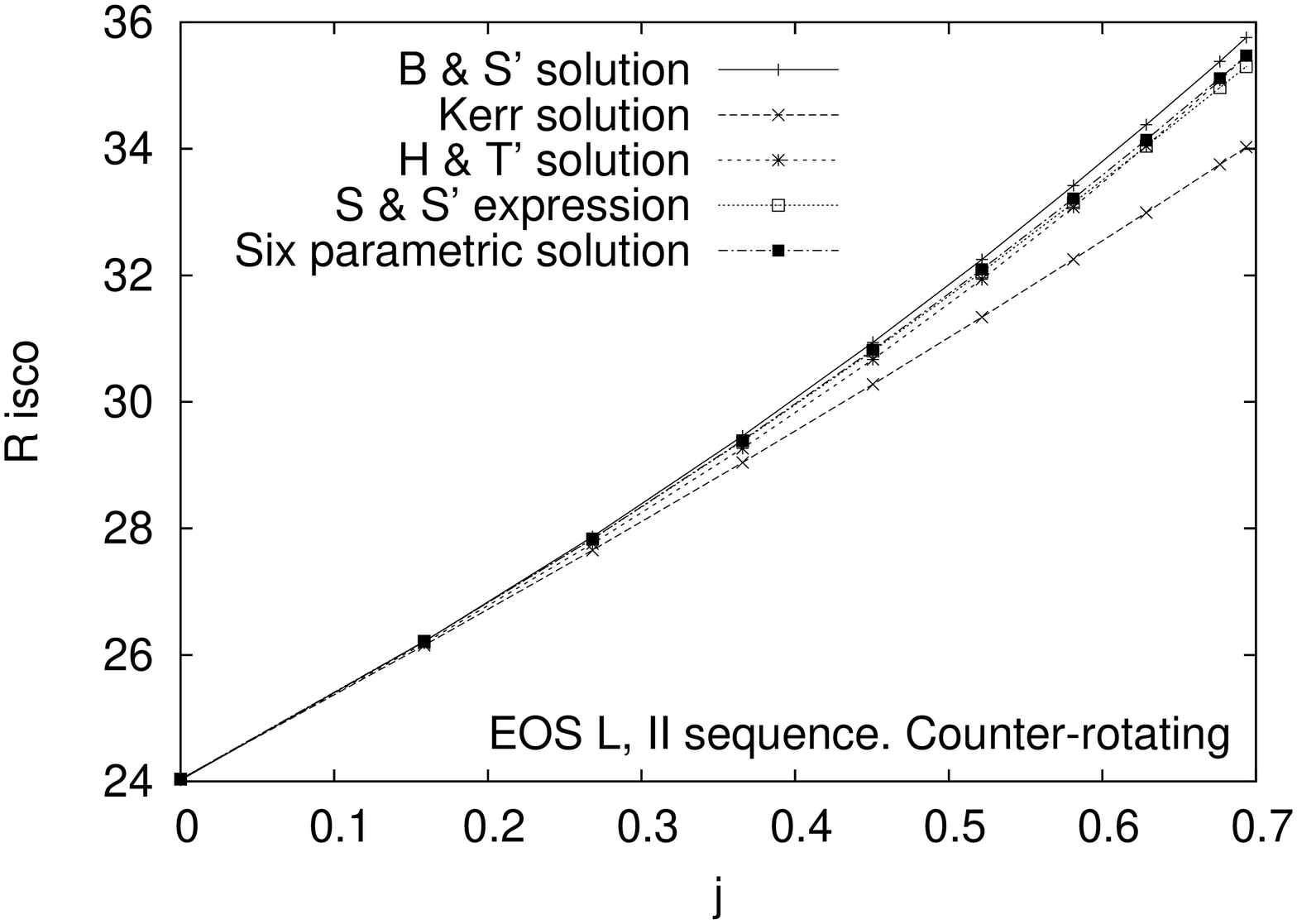}&
\hspace{-0.5cm}\includegraphics[width=8.8cm,height=8.4cm]{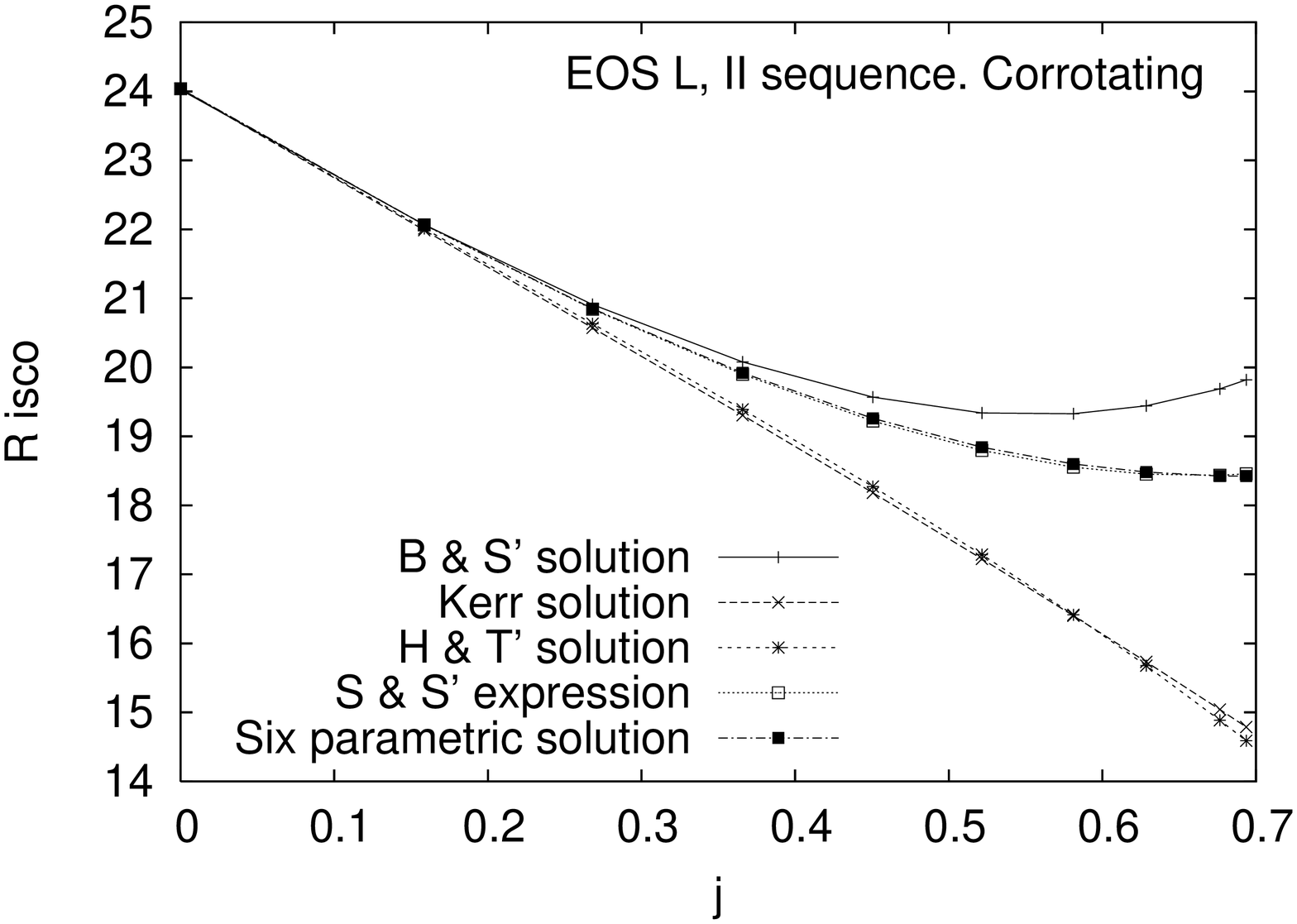}\\
\vspace{-0.5cm} (e.1) & (e.2)\\
\hspace{-0.5cm}\includegraphics[width=8.8cm,height=8.4cm]{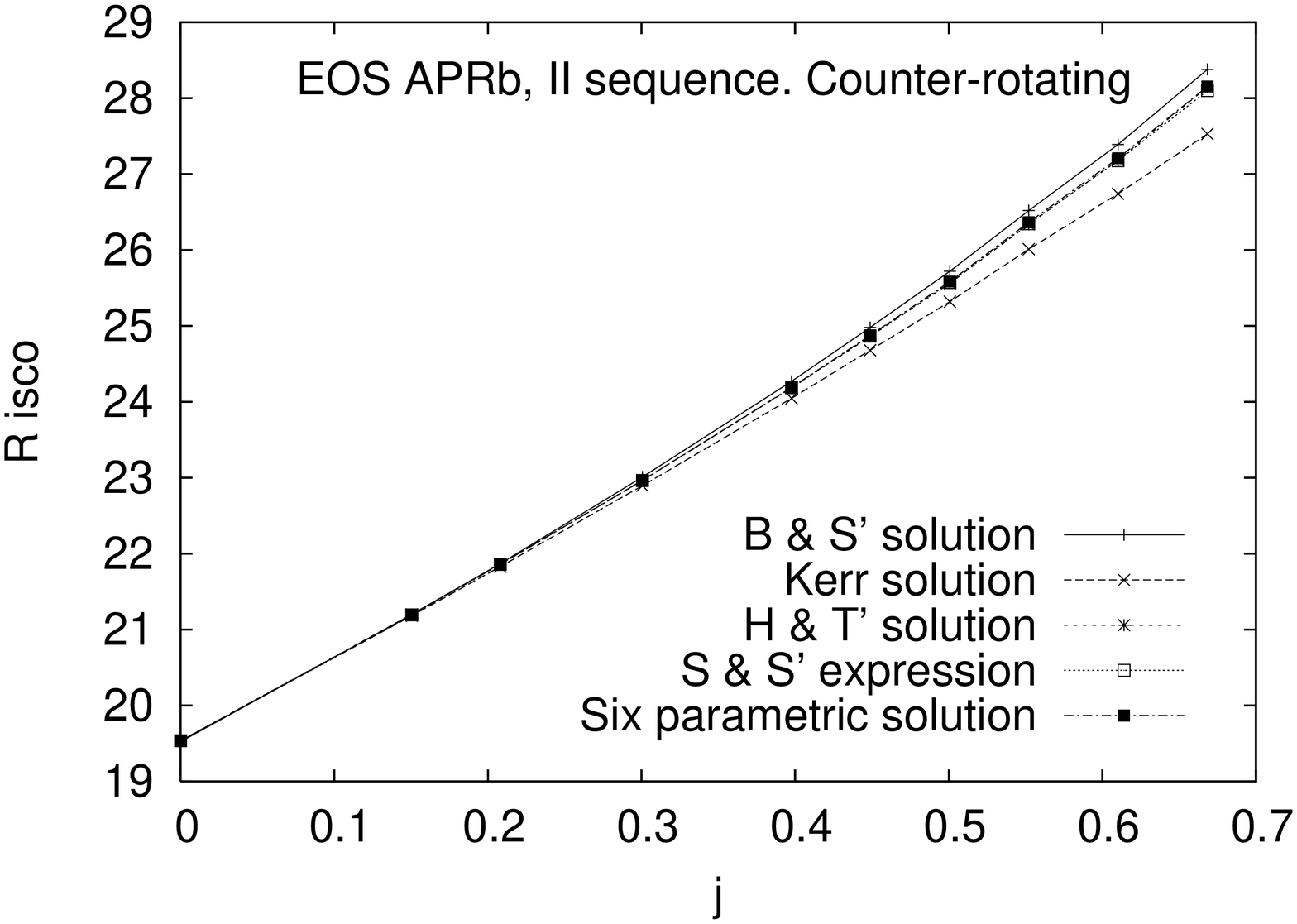}&
\hspace{-0.5cm}\includegraphics[width=8.8cm,height=8.4cm]{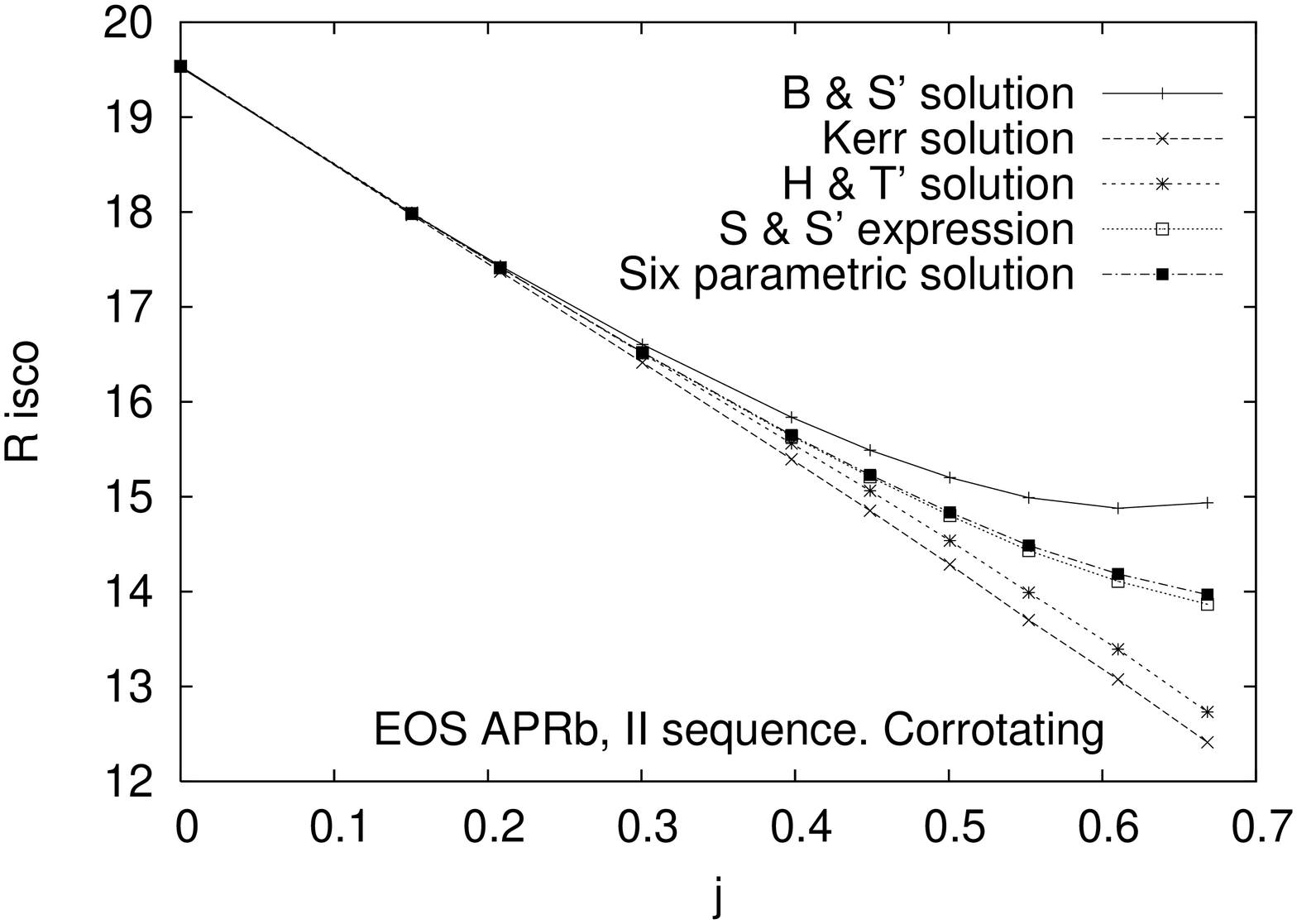}
\end{tabular}
\vspace{-0.5cm}\caption{ISCO for the EOS A sequence with constant rest mass corresponding to a
maximum-mass model in the non rotating limit of 1.948$M_\odot$ for (a.1) the counter--rotating
orbits and (a.2) the co--rotating orbits. ISCO for the EOS AU sequence with constant rest mass
corresponding to a non–rotating model of 1.578$M_\odot$ for (b.1) the counter--rotating orbits and
(b.2) the co--rotating orbits. ISCO for EOS FPS sequence with constant rest mass corresponding to a
maximum-mass model in the non rotating limit of 2.226$M_\odot$ for (c.1) the counter--rotating
orbits and (c.2) the co--rotating orbits. ISCO for EOS L sequence with constant rest mass
corresponding to the maximum-mass model in the non rotating limit 3.232$M_\odot$ for (d.1) the
counter--rotating orbits and (d.2) the co--rotating orbits. ISCO for EOS APRb sequence with
constant rest mass corresponding to the maximum-mass model in the non rotating limit 2.672$M_\odot$
for (e.1) the counter--rotating o
 rbits and (e.2) the co--rotating orbits.}\label{Fig:AjusteROME1}
\end{figure}
\end{center}
\end{widetext}

We can see that our solution presents a better adjust for the radii of
the ISCO in comparison with the standard analytic models used in the literature to describe the
exterior gravitational field of a rotating neutron star, e.g. Kerr solution, Hartle \& Thorne'
solution (H \& T' solution) and the  Shibata \& Sasaki' expressions (S \& S' expression). In
average, for all cases the accuracy of  $R_+$ is better than 2\%, and for $R_-$ better than
1\%. The fact that the six parametric solution fits quite well from the non-rotating limit to
the rapidly rotations rates, in all sequences of mass in the five studied equation of state,
implies that it could be an appropriated analytic closed form model for the exterior gravitational
field of a neutron star.

\section{Concluding Remarks}
\label{sec:Conclusions}%%%%%

We present a new stationary axisymmetric six--parameter closed--form analytic solution generalizing to
Kerr--Newmann solution with arbitrary mass--quadrupole moment, octupole current moment and electric charge.
The analytic form of its multipolar structure and their electric and magnetic fields, are also presented.

We calculated some properties of the solution, using realistic data of interior numerical solutions, in
order to demonstrate that the solution could model the exterior field of a realistic Neutron Star.  In the
electrovacuum case, we used the magnetic dipolar moment given by Bocquet \emph{et al.} in
\cite{BonazzolaBocquet}; in the vacuum case, we matched it to highly--accurate numerical solutions by
Berti \& Stergioulas \cite{BertiStergioulas}, imposing that the mass, angular momentum, mass--quadrupole and current
octupole moment of the numerical and analytic space--times have the same value. We also showed that this
six--parameter analytic closed--form model is an appropriated model for the exterior field of a
slowly or rapidly Neutron Star based on comparisons of the radii of ISCO obtained with a) Berti \&
Stergioulas numerical solutions (2004) \cite{BertiStergioulas}, b) Kerr solution (1963)
\cite{Kerr}, c) Hartle \& Thorne solution \cite{HartleThorne}, d) an analytic series expansion
derived by Shibata \& Sasaki (1998) \cite{Shibata} with our analytic model for all Equations of
State given in \cite{BertiStergioulas}. However, the solution's accuracy should be tested also
outside of the equatorial plane, and through the calculation of other physical observables, which
is a task for the future.

The exact solution could be used in future studies of astrophysical plasma, dynamic of geodesics and accretion
of particles in the surrounding space--time of the Neutron Star.

\section*{Acknowledgments}
J. D. Sanabria-G\'{o}mez acknowledges financial support from COLCIENCIAS Colombia.  Jorge A. Rueda
thanks Universidad Industrial de Santander (Colombia) and Universidad de Los Andes (Venezuela) for
the financial support.  This work was also supported by Project 5116 (DIEF--Ciencias of the
Universidad Industrial de Santander, Colombia). The authors thank to referee for helpful comments.

%%%%%%%%%%%%%%%%%%%%%%%%%%%

\suppressfloats
\end{document}